\documentclass[%
 reprint,
 superscriptaddress,
 amsmath,
 amssymb,
 amsfonts,
 aps,
 prab,
]{revtex4-2}

\usepackage{graphicx}
\usepackage{dcolumn}
\usepackage{bm}
\usepackage{hyperref}
\usepackage{mathtools}
\usepackage{upgreek}

\begin{document}

\title{Bayesian optimization of electron energy from laser wakefield accelerators}

\author{P.~Valenta}
\email{petr.valenta@eli-beams.eu}
\affiliation{ELI Beamlines Facility, The Extreme Light Infrastructure ERIC, Za Radnicí 835, 25241 Dolní Břežany, Czech Republic}

\author{T.~Zh.~Esirkepov}
\affiliation{Kansai Institute for Photon Science, National Institutes for Quantum Science and Technology, Umemidai 8-1-7, Kizugawa, Kyoto 619-0215, Japan}

\author{J.~D.~Ludwig}
\affiliation{Lawrence Livermore National Laboratory, 7000 East Avenue, Livermore, California 94551,
USA}

\author{S.~C.~Wilks}
\affiliation{Lawrence Livermore National Laboratory, 7000 East Avenue, Livermore, California 94551,
USA}

\author{S.~V.~Bulanov}
\affiliation{ELI Beamlines Facility, The Extreme Light Infrastructure ERIC, Za Radnicí 835, 25241 Dolní Břežany, Czech Republic}

\date{\today}

\begin{abstract}

We use Bayesian optimization in combination with three-dimensional particle-in-cell simulations to determine the optimal laser and plasma parameters that, for a given laser pulse energy, maximize the cut-off energy of an electron beam produced by laser wakefield accelerators. We assume a Gaussian laser driver with matched spot size and amplitude and investigate both self-guiding in a uniform-density plasma and guiding in a preformed plasma channel with matched radius. To interpret the simulation results quantitatively, we derive novel analytical expressions for the maximum electron energy and the corresponding acceleration distance, accounting for the effects of laser diffraction and energy depletion. Based on the results obtained, we discuss the potential scalability of the optimal input (plasma density, pulse duration, amplitude, spot size, and channel radius) and output (electron energy, electron charge, acceleration length, and acceleration efficiency) parameters to laser systems of arbitrary energy.

\end{abstract}

\maketitle

\section{Introduction \label{sec:1}}

Laser wakefield acceleration (LWFA) is a rapidly evolving technique that stands out for its ability to achieve acceleration gradients several orders of magnitude higher than those of conventional accelerators \cite{tajima1979, esarey2009a, gonsalves2019a}. This method relies on the interaction between an intense laser pulse and plasma, where the laser drives a plasma wave that can trap and accelerate electrons to relativistic energies. Over the past few decades, LWFA has gained significant attention for its potential to produce compact electron accelerators with a wide range of applications, from advanced radiation sources \cite{gruner2007, corde2013a, bulanov2013a, albert2014, kurz2021} to the exploration of strong-field quantum electrodynamics phenomena \cite{gonoskov2022, yu2024}.

In LWFA, proper selection of initial laser and plasma parameters is crucial for optimizing various characteristics of the accelerated electron beam. Such optimization, however, presents a significant challenge due to the highly nonlinear and multi-parametric nature of the problem. Furthermore, it is often necessary to optimize multiple, potentially conflicting objectives simultaneously, which requires identifying an appropriate trade-off.

Analytical models offer insight into how certain electron beam properties depend on input parameters, though they inevitably rest on simplifying assumptions. In practice, the effects of the self-consistent evolution of the laser and plasma significantly limit the applicability of such models, and alternative methods may be required. Numerical simulations are often employed to account for the complex interactions between the laser and plasma, yet simulations with sufficient spatial and temporal resolution to accurately capture the relevant physics are computationally very expensive. On the other hand, the range of parameters that can be explored through experimental methods is intrinsically limited by the capabilities of existing laser systems, optics, and diagnostics.

In this work, we aim to identify the optimal set of laser and plasma parameters that, for a given laser pulse energy, maximize the cut-off energy (i.e., the energy threshold beyond which there are very few or no particles) of an electron beam accelerated via LWFA. At this stage, we focus solely on optimizing the cut-off energy, without regard to other electron beam parameters. This focus is motivated by the needs of specific applications where achieving a cut-off energy above certain threshold is essential (e.g., muon production \cite{terzani2025, calvin2025, zhang2025, ludwig2025} and nuclear activation \cite{nedorezov2021, kolenaty2022}).

To reduce the number of free parameters, we make the following assumptions: (i) LWFA is driven by a linearly polarized Gaussian laser pulse with a constant spectral phase; (ii) the plasma has a uniform density profile along the laser propagation direction; (iii) LWFA is operated in the matched regime \cite{lu2006}; and (iv) electrons are injected using the nanoparticle-assisted method \cite{cho2018}. Although we keep the key parameters free, it is important to note that under these assumptions, the maximum possible electron energy is not necessarily achieved (see, e.g., LWFA using non-Gaussian \cite{beaurepaire2015, oumbarekespinos2023} or chirped \cite{kalmykov2012, kim2017a} laser pulses, tailored plasma targets \cite{sprangle2001, guillaume2015}, or the strongly mismatched regime \cite{perevalov2020, poder2024}). More advanced configurations, however, introduce additional parameters, further complicating the optimization process. We anticipate considering such cases in future work.

To achieve our goal, we use Bayesian optimization (BO), where the objective function (i.e., the observable to be optimized -- in our case, the electron cut-off energy) is sampled using three-dimensional (3D) particle-in-cell (PIC) simulations. Our task thus involves optimizing an expensive-to-evaluate, derivative-free, and potentially non-convex function. For such problems, BO is particularly well suited. BO, including its modifications (e.g., multi-objective and multi-fidelity BO), has already been successfully applied to address complex tasks in LWFA \cite{shalloo2020, jalas2021, kirchen2021, jalas2023, ferranpousa2023, irshad2023, irshad2024} and accelerator physics in general \cite{roussel2024, roussel2024a}.

We demonstrate our approach using a laser pulse with relatively low energy ($ 10 \, \mathrm{mJ} $), as in this case the optimal acceleration distance is reasonably long (less than $ 500 \, \upmu\mathrm{m} $ in all simulated cases). This makes the set of PIC simulations required for the convergence of BO computationally feasible. We optimize two scenarios -- LWFA with and without a preformed plasma channel of matched radius -- to investigate its impact on the electron energy. To interpret the simulation results quantitatively, we derive novel analytical expressions for predicting the maximum electron energy and the corresponding acceleration distance, taking into account the effects of laser diffraction and energy depletion. Additionally, we express the optimization results using dimensionless parameters and discuss their potential scalability to lasers of arbitrary energy.

The remainder of this paper is organized as follows. In Sec.~\ref{sec:2}, we revisit the fundamental theory of relativistically strong laser pulse propagation in tenuous plasma. In Sec.~\ref{sec:3}, we introduce novel analytical expressions for the LWFA electron energy and acceleration length, considering both unguided (Sec.~\ref{sec:3a}) and guided (Sec.~\ref{sec:3b}) laser pulses. Sections~\ref{sec:4} and \ref{sec:5} describe the setup of the BO algorithm and the PIC simulations, respectively. In Sec.~\ref{sec:6}, we present the optimization results for the cases of self-guiding in a uniform-density plasma (Sec.~\ref{sec:6a}) and guiding in a preformed plasma channel with matched radius (Sec.~\ref{sec:6b}). We discuss the potential scalability of the optimal parameters in Sec.~\ref{sec:7}, and finally, we summarize our findings in Sec.~\ref{sec:8}.

\section{Propagation of intense laser pulse in plasma \label{sec:2}}

The profile of a Gaussian laser pulse can be characterized by its vector potential,
\begin{equation}\label{eq:gauss}
    A = \tilde{a} \exp{\left( -i \psi \right)}.
\end{equation}
Here, a scalar description is used (i.e., ignoring polarization). In what follows, we assume that the vector potential is normalized by $ m_e c / e $, where $ m_e $ is the electron mass, $ c $ is the speed of light in vacuum, and $ e $ is the elementary charge.

For a radially symmetric profile, the amplitude $ \tilde{a} (x, r_{\perp}, \xi) $ and phase $ \psi (x, r_{\perp}, t) $ depend on the longitudinal coordinate $ x $, radial coordinate $ r_{\perp} $, time $ t $, and the co-moving coordinate $ \xi = x - \beta c t $, which follows the laser pulse at the group velocity $ \beta c $, as
\begin{equation}\label{eq:a_tilde}
    \tilde{a} = a \exp{\left(-4 \ln{2} \frac{\xi^2}{l_0^2} - \frac{r_{\perp}^2}{w^2} \right)},
\end{equation}
and
\begin{equation}\label{eq:psi}
    \psi = \omega_0 t - k_x x \left(1 - \frac{r_{\perp}^2}{2 \left( x^2 + x_R^2 \right)} \right) - \arctan\left(\frac{x}{x_R}\right),
\end{equation}
respectively. Here, $ \omega_0 $, $ k_x = \omega_0 / c $, and $ l_0 = \sqrt{2} c \beta \tau_0 $ denote the angular frequency, wavenumber, and the finite length of the laser pulse, respectively, where $ \tau_0 $ is the full-width-at-half-maximum (FWHM) duration measured from the intensity profile.

The functions $ a (x) $ and $ w (x) $ in Eq.~(\ref{eq:a_tilde}) are defined as
\begin{equation}\label{eq:a}
    a = a_0 \frac{w_0}{w}
\end{equation}
and
\begin{equation}\label{eq:w}
    w = w_0 \sqrt{1 + \frac{x^2}{x_R^2}},
\end{equation}
where $ a_0 = e E_0 / m_e \omega_0 c $ is the strength parameter of the laser pulse (with $ E_0 $ being the electric field amplitude) and $ w_0 $ is the laser waist (i.e., the radius at which the intensity drops to $ 1 / \mathrm{e}^2 $ of its peak value at the focus). The Rayleigh length, $ x_R = k_x w_0^2 / 2 $, is defined as the distance from the focus along the propagation direction to the point where the laser radius increases by a factor of $ \sqrt{2} $ compared to the waist.

Further, we assume that the laser central wavelength, $ \lambda_0 = 2 \pi c / \omega_0 $, and energy,
\begin{equation}\label{eq:ene_0}
    \mathcal{E}_0 = \frac{\pi^{3/2}}{16 \sqrt{\ln{2}}} \overline{\mathcal{E}} a_0^2 \frac{w_0^2}{\lambda_0^2} \omega_0 \tau_0,
\end{equation}
are given. Here, 
\begin{equation}\label{eq:e_bar_p_bar}
    \overline{\mathcal{E}} = \frac{\pi \overline{\mathcal{P}}}{\omega_0} \quad \mathrm{and} \quad \overline{\mathcal{P}} = \frac{2 m_e c^3}{r_e},
\end{equation}
$ r_e = e^2 / 4 \pi \epsilon_0 m_e c^2 $ is the classical electron radius, and $ \epsilon_0 $ is the vacuum permittivity. Note that $ \overline{\mathcal{E}} \approx 29 \ \mathrm{\upmu J} $ for $ \lambda_0 = 1 \ \mathrm{\upmu m} $ and $ \overline{\mathcal{P}} \approx 17.4 \ \mathrm{GW} $. The laser power is then
\begin{equation}\label{eq:p_0}
    \mathcal{P}_0 = 2 \sqrt{\frac{\ln{2}}{\pi}} \frac{\mathcal{E}_0}{\tau_0}.
\end{equation}

Upon propagating through a fully ionized plasma, a relativistically strong laser pulse (i.e., $ a_0 \gtrsim 1 $) can either be guided by some form of external optical guiding or become self-guided. Whether the pulse is self-guided depends on the relationship between its power and the critical power for relativistic self-focusing \cite{sun1987},
\begin{equation}\label{eq:p_cr}
    \mathcal{P}_{\mathrm{cr}} = \overline{\mathcal{P}} \frac{n_{\mathrm{cr}}}{n_e},
\end{equation}
where $ n_e $ is the electron density and $ n_{\mathrm{cr}} = \pi / r_e \lambda_0^2 $ is the critical plasma density.

When $ \mathcal{P}_0 < \mathcal{P}_{\mathrm{cr}} $, the laser pulse is not self-guided, meaning it diffracts as it propagates through the plasma. Diffraction reduces the laser amplitude, preventing electrons from being efficiently accelerated over a sufficiently long distance. On the other hand, when $ \mathcal{P}_0 \gg \mathcal{P}_{\mathrm{cr}} $, nonlinear effects (e.g., laser filamentation instability \cite{naumova2002, valenta2021a}, vortex generation \cite{bulanov1996}, and soliton formation \cite{bulanov1999, esirkepov2002}) may arise, distorting the laser pulse profile and hindering the acceleration process. Therefore, in the case of self-guiding, maintaining an appropriate balance between the laser power and the critical power for relativistic self-focusing is crucial to ensure effective electron acceleration in plasma. We express the ratio of $ \mathcal{P}_0 $ to $ \mathcal{P}_{\mathrm{cr}} $ as a function of $ n_e $ and $ \tau_0 $,
\begin{equation}\label{eq:p_0_over_p_cr}
\begin{aligned}
    \frac{\mathcal{P}_0}{\mathcal{P}_{\mathrm{cr}}} & = \frac{2 \sqrt{\ln{2}} \, r_e}{\pi^{3/2} \overline{\mathcal{P}}} \frac{\lambda_0^2 \mathcal{E}_0 n_e}{\tau_0} \\
    & \approx 4.84 \times 10^{-26} \, \mathrm{\frac{m \, s}{J}} \, \frac{\lambda_0^2 \mathcal{E}_0 n_e}{\tau_0}.
\end{aligned}
\end{equation}

To maximize the amplitude of the accelerating field, the pulse duration must be resonant with the plasma density. This condition is determined by the specific longitudinal pulse shape. For a Gaussian profile and $ a_0 \ll 1 $, the resonant duration can be calculated analytically as $ \tau_r = 2 \sqrt{2 \ln{2}} / \omega_p $ \cite{leemans1996}, where $ \omega_p = \sqrt{n_e e^2 / m_e \epsilon_0} $ is the plasma frequency. The ratio of $ \tau_0 $ to $ \tau_r $ can be expressed in terms of $ n_e $ and $ \tau_0 $ as
\begin{equation}\label{eq:tau_0_over_tau_r}
    \frac{\tau_0}{\tau_r} = \frac{\tau_0 \omega_p}{2 \sqrt{2 \ln{2}}} \approx 23.96 \, \left( \mathrm{\frac{m^3}{s^2}} \right)^{1/2} \, \left( \tau_0^2 n_e \right)^{1/2}.
\end{equation}

For self-guided propagation of a laser pulse in plasma without significant changes to the pulse profile over multiple Rayleigh lengths, the laser strength parameter and waist should be set according to the matching conditions \cite{lu2006}. Here, we rewrite the matching conditions in terms of $ n_e $ and $ \tau_0 $ as
\begin{equation}\label{eq:a_0}
\begin{aligned}
    a_0 & = 2 \left( \frac{\mathcal{P}_0}{\mathcal{P}_{\mathrm{cr}}} \right)^{1/3} \\
        & \approx 7.29 \times 10^{-9} \, \left( \mathrm{\frac{m \, s}{J}} \right)^{1/3} \, \left( \frac{\lambda_0^2 \mathcal{E}_0 n_e}{\tau_0} \right)^{1/3}
\end{aligned}
\end{equation}
and
\begin{equation}\label{eq:w_0}
\begin{aligned}
    w_0 & = \frac{2 c \sqrt{a_0}}{\omega_p} \\
        & \approx 9.07 \times 10^2 \, \left( \mathrm{\frac{s}{J \, m^2}} \right)^{1/6} \, \left( \frac{\lambda_0^2 \mathcal{E}_0}{n_e^2 \tau_0} \right)^{1/6}.
\end{aligned}
\end{equation}

In the case of laser guiding in a preformed plasma channel, we assume that the electron density profile has a parabolic dependence on the radial coordinate \cite{bobrova2001}:
\begin{equation}\label{eq:plasma_channel}
    n_e\left(r_{\perp}\right) \approx n_e\left(0\right) \left( 1 + 0.33 \frac{r_{\perp}^2}{R_{\mathrm{ch}}^2} \right).
\end{equation}
Here, $ R_{\mathrm{ch}} $ is the matched radius of the channel, which can be estimated as \cite{bobrova2001}
\begin{equation}\label{eq:matched_radius}
    R_{\mathrm{ch}} \approx 0.57 w_0^2 \sqrt{\pi r_e n_e\left(0\right)}.
\end{equation}

Equations~(\ref{eq:p_0_over_p_cr})--(\ref{eq:matched_radius}) suggest that the problem of maximizing the LWFA electron energy, under the considerations specified above, reduces to finding an optimal combination of only two parameters: $ n_e $ and $ \tau_0 $ (or, equivalently, $ \mathcal{P}_0 $).

\section{electron acceleration \label{sec:3}}

The maximum energy that can be imparted to an electron through LWFA is determined by the strength of the accelerating field and the length over which acceleration occurs, both of which depend on multiple interrelated factors. In addition to the diffraction of unguided laser pulses mentioned in Sec.~\ref{sec:2}, electron energy may be limited by dephasing (i.e., electrons outrunning the accelerating phase of the wakefield) or pump depletion (i.e., the laser losing energy while propagating through plasma and driving a wakefield).

Both diffraction and pump depletion lead to modifications of the laser pulse shape. To account for these effects, we redefine the functions $ a (x) $ and $ w (x) $ given by Eqs.~(\ref{eq:a}) and (\ref{eq:w}), respectively, as follows \cite{esarey2009a}:
\begin{equation}\label{eq:a_plasma}
    a = a_0 \frac{w_0}{w} \exp{\left( - \frac{x}{l_{pd}} \right)},
\end{equation}
\begin{equation}\label{eq:w_plasma}
    w = w_0 \sqrt{1 + \frac{x^2}{x_R^2} \left( 1 - \frac{\mathcal{P}_0}{\mathcal{P}_{\mathrm{cr}}} \right)}.
\end{equation}
Here, $ l_{pd} $ is the pump depletion length (i.e., the characteristic scale over which the laser deposits energy into the plasma wave). For simplicity, Eq.~(\ref{eq:a_plasma}) models pump depletion by assuming an exponential decay of the laser amplitude over the distance $ l_{pd} $. Equation~(\ref{eq:w_plasma}) predicts ``catastrophic'' focusing when $ \mathcal{P}_0 / \mathcal{P}_{\mathrm{cr}} > 1 $; in reality, however, this is mitigated by higher-order effects \cite{sprangle1987, hafizi2000}.

The longitudinal electric field generated by a laser pulse in plasma can be calculated analytically within the framework of the linear approximation \cite{esarey2009a, terzani2023}. In units of $ m_e \omega_p c / e $, the electric field is given by
\begin{equation}\label{eq:ex}
    E_x = \frac{\sqrt{\pi} a^2 k_p l_0}{8 \sqrt{2 \ln{2}}} \exp{\left(-\frac{k_p^2 l_0^2}{32 \ln{2}} - \frac{2 r_{\perp}^2}{w^2} \right)} \cos{\left( k_p \xi + \phi \right)},
\end{equation}
where $ k_p = \omega_p / c $ is the plasma wavenumber and $ \phi $ is the initial phase of the plasma wave. The net change in the kinetic energy of an ultra-relativistic electron accelerated along the longitudinal coordinate by the electric field given in Eq.~(\ref{eq:ex}) can be calculated as
\begin{equation}\label{eq:ene}
    \frac{\Delta \mathcal{E}_e}{m_e c^2} = k_p \int_{0}^{+\infty} E_x \left[x, r_{\perp} = 0, \xi \approx x \left(1 - \beta_{\mathrm{ph}}\right) \right] \mathrm{d}x,
\end{equation}
where $ \beta_{\mathrm{ph}} \approx \beta \approx \sqrt{1 - n_e / n_{\mathrm{cr}}} $ is the phase velocity of the plasma wave normalized by $ c $.

\subsection{Unguided pulse \label{sec:3a}}

For an unguided laser pulse (i.e., $ \mathcal{P}_0 / \mathcal{P}_{\mathrm{cr}} \rightarrow 0 $), the effect of laser diffraction is typically much more detrimental than pump depletion. Neglecting pump depletion (i.e., $ l_{pd} \rightarrow +\infty $) and evaluating the integral in Eq.~(\ref{eq:ene}), we obtain
\begin{equation}\label{eq:ene_unguided}
\begin{aligned}
    \frac{\Delta \mathcal{E}_e}{m_e c^2} &= \frac{\pi^{3/2}}{8 \sqrt{2 \ln{2}}} a_0^2 k_p^2 l_0 x_R \cos{\phi} \\
    &\times \exp{\left[ - \frac{k_p^2 l_0^2}{32 \ln{2}} - k_p x_R \left(1 - \beta_{\mathrm{ph}}\right) \right]}.
\end{aligned}
\end{equation}
Expression~(\ref{eq:ene_unguided}) reaches its maximum for $ k_p l_0 = 4 \sqrt{\ln{2}} $ (i.e., $ \tau_0 = \tau_r $) and $ \phi = 0 $,
\begin{equation}\label{eq:max_ene_unguided}
    \frac{\mathcal{E}_{e, \mathrm{max}}}{m_e c^2} = \frac{\pi^{3/2}}{2 \sqrt{2}} a_0^2 k_p x_R \exp{\left[ - k_p x_R \left(1 - \beta_{\mathrm{ph}} \right) - \frac{1}{2} \right]}.
\end{equation}

Considering a laser pulse with spot size given by Eq.~(\ref{eq:w_0}), the Rayleigh length is $ x_R \approx 2 \gamma_{\mathrm{ph}} a_0 / k_p $. Using the approximation $ 1 - \beta_{\mathrm{ph}} \approx 1 / 2 \gamma_{\mathrm{ph}}^2 $, where $ \gamma_{\mathrm{ph}} = (1 - \beta_{\mathrm{ph}}^2)^{-1/2} $ is the Lorentz factor of the plasma wave (i.e., the low-density plasma limit), the expression for the maximum electron energy in the case of an unguided laser pulse simplifies to
\begin{equation}\label{eq:max_ene_unguided_2}
    \frac{\mathcal{E}_{e, \mathrm{max}}}{m_e c^2} \approx 2.39 \, a_0^3 \gamma_{\mathrm{ph}} \exp{\left(-\frac{a_0}{\gamma_{\mathrm{ph}}} \right)}.
\end{equation}
The corresponding acceleration length, which is primarily limited by laser diffraction, is given by
\begin{equation}\label{eq:l_acc_unguided}
    \frac{l_{\mathrm{acc}}}{\lambda_0} \approx 0.32 \, a_0 \gamma_{\mathrm{ph}}^2.
\end{equation}

\subsection{Guided pulse \label{sec:3b}}

For a guided laser pulse (i.e., $ \mathcal{P}_0 / \mathcal{P}_{\mathrm{cr}} \rightarrow 1 $), diffraction is suppressed, and the dominant mechanism limiting the maximum electron energy becomes pump depletion. By evaluating the integral in Eq.~(\ref{eq:ene}) for the guided case, we obtain
\begin{equation}
\begin{aligned}\label{eq:ene_guided}
    \frac{\Delta \mathcal{E}_e}{m_e c^2} &= \frac{\sqrt{\pi}}{8 \sqrt{2 \ln{2}}} a_0^2 k_p^2 l_0 l_{pd} \exp{\left( - \frac{k_p^2 l_0^2}{32 \ln{2}} \right)} \\
    &\times \frac{2 \cos{\phi} - k_p l_{pd} \left( 1 - \beta_{\mathrm{ph}} \right) \sin{\phi}}{4 + k_p^2 l_{pd}^2 \left( 1 - \beta_{\mathrm{ph}} \right)^2}.
\end{aligned}
\end{equation}

Taking into account the pump depletion length in the form $ l_{pd} \approx l_0 k_0^2 / k_p^2 $ \cite{lu2007}, Eq.~(\ref{eq:ene_guided}) reaches its maximum for $ k_p l_0 \approx 4.07 $ (i.e., $ \tau_0 > \tau_r $) and $ \phi \approx -0.79 $. Assuming further that $ \gamma_{\mathrm{ph}} \gg 1 $, the expression for the maximum electron energy in the case of a guided laser pulse simplifies to
\begin{equation}\label{eq:max_ene_guided}
    \frac{\mathcal{E}_{e, \mathrm{max}}}{m_e c^2} \approx 0.52 \, a_0^2 \gamma_{\mathrm{ph}}^2.
\end{equation}
The corresponding acceleration length, which is primarily limited by pump depletion, is given by
\begin{equation}\label{eq:l_acc_guided}
    \frac{l_{\mathrm{acc}}}{\lambda_0} \approx 0.23 \, \omega_p \tau_0 \gamma_{\mathrm{ph}}^3.
\end{equation}

\section{Bayesian optimization \label{sec:4}}

The key components of the BO method are the probabilistic surrogate model of the objective function, which fits the observed data points while quantifying uncertainty in unobserved regions, and the acquisition function, which determines the next point to evaluate. We employ a Gaussian process (GP) \cite{rasmussen2005} as the surrogate model and the upper confidence bound (UCB) \cite{auer2002} as the acquisition function.

Kernels (also referred to as covariance functions) determine the shape of both the prior and posterior distributions of the GP. We use the Matérn kernel, which provides a good balance between smoothness and flexibility for modeling non-linear objective landscapes typical of LWFA optimization. The Matérn kernel includes a parameter $ \nu $ that controls the smoothness of the resulting function. We chose $ \nu = 5/2 $, meaning the learning function is (at least) twice differentiable.

A critical aspect of BO is balancing exploration (i.e., searching the broader parameter space) and exploitation (i.e., probing areas near the current known optimum). The UCB acquisition function includes a free parameter $ \kappa $ that allows us to control this balance. We used a conservatively decaying $ \kappa $ \cite{srinivas2012} to modestly shift the search toward exploitation in later iterations, while preserving some exploration capability given the limited evaluation budget.

Based on the reasoning presented in Sec.~\ref{sec:2}, we initiate the search for the optimal case near the intersection of the contour curves $ \mathcal{P}_0 = \mathcal{P}_{\mathrm{cr}} $ and $ \tau_0 = \tau_r $. For a laser with $ \mathcal{E}_0 = 10 \, \mathrm{mJ} $ and $ \lambda_0 = 1 \, \upmu\mathrm{m} $, these curves, given by
\begin{equation}\label{eq:p_0_over_p_cr_contour}
    n_e \approx 2.07 \times 10^{39} \, \mathrm{m^{-3} \, s^{-1}} \, \tau_0
\end{equation}
and
\begin{equation}\label{eq:tau_0_over_tau_r_contour}
    \tau_0 \approx 4.17 \times 10^{-2} \, \mathrm{m^{-3/2} \, s} \, n_e^{-1/2},
\end{equation}
respectively, intersect at $ n_e \approx 1.95 \times 10^{19} \, \mathrm{cm^{-3}} $ and $ \tau_0 \approx 9.44 \, \mathrm{fs} $. At the intersection, the matching conditions (\ref{eq:a_0}) and (\ref{eq:w_0}) yield $ a_0 = 2 $ and $ w_0 \approx 3.4 \, \upmu\mathrm{m} $, respectively. We restrict the parameter space for the BO algorithm to $ \tau_0 \in [ 3, 12 ] \, \mathrm{fs} $ and $ n_e \in [ 1, 4 ] \times 10^{19} \, \mathrm{cm^{-3}} $. If these bounds prove inadequate (i.e., when the current optimum lies on or very close to the search space boundary), they can be dynamically adjusted during the optimization process, allowing the optimizer to explore the surrounding region.

\begin{figure*}[t]
\includegraphics[width=1.0\linewidth]{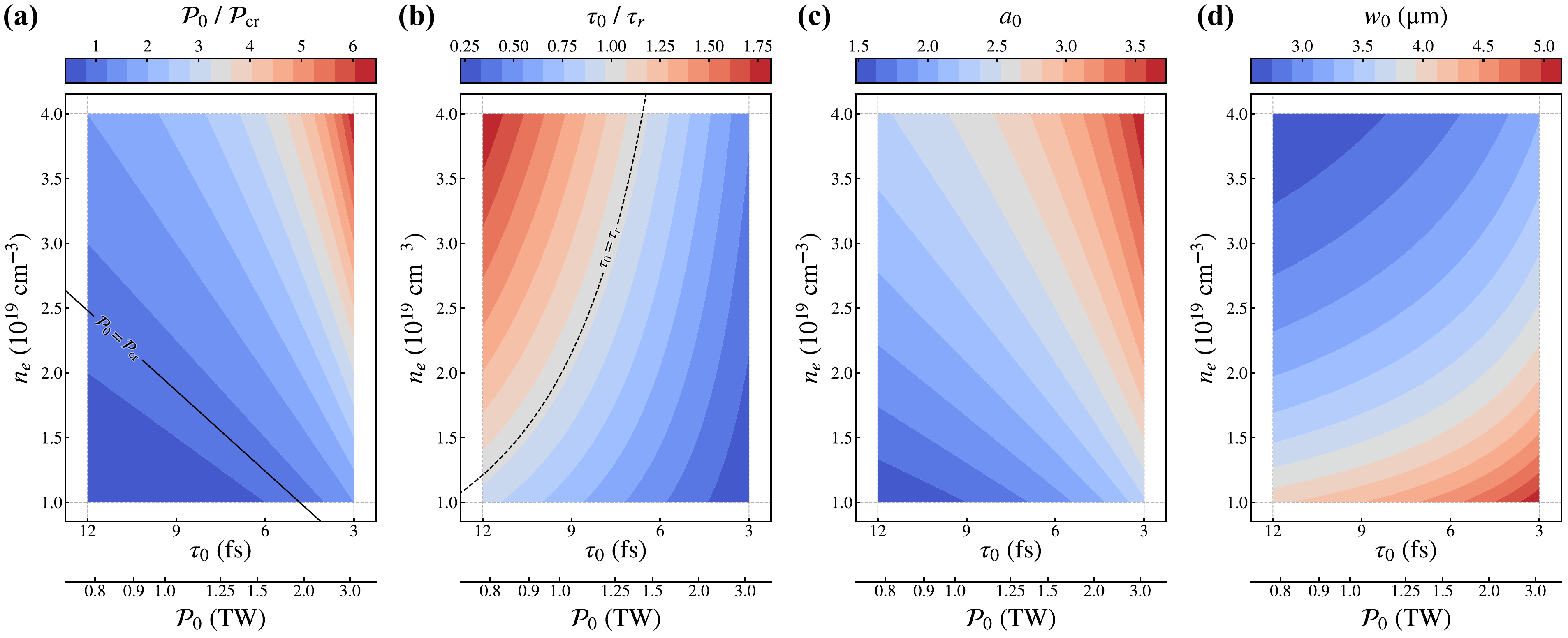}
\caption{(Color online). (a) Ratio of laser power to critical power for relativistic self-focusing, $ \mathcal{P}_0 / \mathcal{P}_{\mathrm{cr}} $, (b) ratio of pulse duration to resonant duration, $ \tau_0 / \tau_r $, (c) strength parameter, $ a_0 $, and (d) beam waist, $ w_0 $, given by Eqs.~(\ref{eq:p_0_over_p_cr})--(\ref{eq:w_0}), respectively. All variables are displayed within the prescribed limits of electron density, $ n_e $, and pulse duration, $ \tau_0 $, (or, equivalently, laser power, $ \mathcal{P}_0 $) considering laser central wavelength $ 1 \, \mathrm{\upmu m} $ and energy $ 10 \, \mathrm{mJ} $. The solid line in (a) and the dashed line in (b) show the contour curves of $ \mathcal{P}_0 = \mathcal{P}_{\mathrm{cr}} $ and $ \tau_0 = \tau_r $ according to Eqs.~(\ref{eq:p_0_over_p_cr_contour}) and (\ref{eq:tau_0_over_tau_r_contour}), respectively.}
\label{fig:1}
\end{figure*}

While we did not implement a specific adaptive stopping criterion (see, e.g., Refs.~\citenum{ishibashi2023} and \citenum{wilson2024}), we monitored the relative improvement in the objective function over the last several evaluations. The optimization was terminated when (i) a sufficiently dense set of evaluated points had been obtained around the current optimum candidate (i.e., the surrounding space was well sampled in all directions) and (ii) additional samples in this region did not significantly affect the estimated value or location of the optimum.

Panels (a)--(d) of Fig.~\ref{fig:1} show $ \mathcal{P}_0 / \mathcal{P}_{\mathrm{cr}} $, $ \tau_0 / \tau_r $, $ a_0 $, and $ w_0 $ according to Eqs.~(\ref{eq:p_0_over_p_cr})--(\ref{eq:w_0}), respectively. All variables are displayed within the prescribed limits of $ n_e $ and $ \tau_0 $, assuming $ \lambda_0 = 1 \, \upmu\mathrm{m} $ and $ \mathcal{E}_0 = 10 \, \mathrm{mJ} $. The contours of $ \mathcal{P}_0 = \mathcal{P}_{\mathrm{cr}} $ and $ \tau_0 = \tau_r $, according to Eqs.~(\ref{eq:p_0_over_p_cr_contour}) and (\ref{eq:tau_0_over_tau_r_contour}), are shown in panels (a) and (b), respectively.

\section{Particle-in-cell simulations \label{sec:5}}

We perform two sets of PIC simulations in 3D Cartesian geometry using the EPOCH code \cite{arber2015}. The laser and plasma parameters used in the simulations are listed in Tabs.~\ref{tab:1} and \ref{tab:2} in the Appendix, corresponding to laser self-guiding in a uniform-density plasma and guiding in a preformed plasma channel with matched radius, respectively. The first four simulations in each case, $ A_1 $ -- $ A_4 $ and $ B_1 $ -- $ B_4 $, are selected based on an educated guess, while the subsequent simulations are determined iteratively by the BO algorithm.

A linearly polarized laser pulse propagates through a fully ionized plasma slab. To suppress artificial plasma wave breaking (and thus electron injection into the plasma wave) at the sharp plasma--vacuum interface, we add a short smooth ramp to the front side of the slab. In the case of a preformed plasma channel, the radial profile of electron density is given by Eq.~(\ref{eq:plasma_channel}). The focal spot of the laser pulse is located $ 10 \, \upmu\mathrm{m} $ from the entrance to the plasma.

A sufficient amount of electron charge is introduced into the plasma wave (while avoiding excessive beam loading \cite{wilks1987}, which would distort the accelerating field) using the method of nanoparticle-assisted injection \cite{cho2018}. A lithium nanoparticle with a radius of $ 30 \, \mathrm{nm} $ is placed on the laser propagation axis, $ 20 \, \upmu\mathrm{m} $ from the entrance to the plasma. At this point, the wakefield structure is already fully developed, but the laser pulse has not yet lost a significant portion of its energy. Additionally, this method ensures that electron injection occurs at the same location in each simulation, regardless of whether self-injection \cite{bulanov1998a} is present, allowing for a fair comparison of electron energies across individual simulation cases. We employ nanoparticle-assisted injection for its relative simplicity; provided that electrons are injected at the same location, the cut-off energy does not depend on the choice of injection mechanism.

The simulations utilize the moving window technique. The simulation window, with longitudinal and transverse dimensions of $ 40 \, \lambda_0 $ and $ 32 \, \lambda_0 $, respectively, moves along the laser propagation direction at a velocity equal to $ c\beta $. The underlying Cartesian grid is uniform, with a resolution of 30 cells per $ \lambda_0 $ in each direction. The simulations are evolved over a time interval sufficient to capture the maximum cut-off energy of the electron beam.

The plasma is cold and collisionless. Electrons are represented by quasi-particles (i.e., computational particles representing a group of physical particles that are close to each other in phase space) with triangular shape functions, whereas ions form a static neutralizing background. Initially, one electron quasi-particle is placed in each grid cell. The equations of motion for quasi-particles are solved using the Boris algorithm \cite{boris1971}, and the electric and magnetic fields are calculated using the standard second-order Yee solver \cite{yee1966}. Absorbing boundary conditions are applied on all sides of the simulation domain for both the fields and quasi-particles.

\section{Optimization results \label{sec:6}}

In what follows, we define the cut-off energy of the accelerated electron beam, $ \mathcal{E}_{e, \mathrm{max}} $, as the upper limit of its energy distribution with a spectral density of at least $ 10 \, \mathrm{fC/MeV} $. The corresponding acceleration length, $ l_{\mathrm{acc}} $, is defined as the distance from injection to the point where the electrons reach their maximum cut-off energy. We do not distinguish whether this length is limited by laser diffraction, energy depletion, electron dephasing, or other effects. A single Bayesian optimization run was performed for each case (self-guiding in uniform-density plasma and guiding in preformed plasma channel with matched radius), and variability across independent runs was therefore not quantified.

\subsection{Self-guiding in uniform-density plasma \label{sec:6a}}

\begin{figure}[t]
\includegraphics[width=1.0\linewidth]{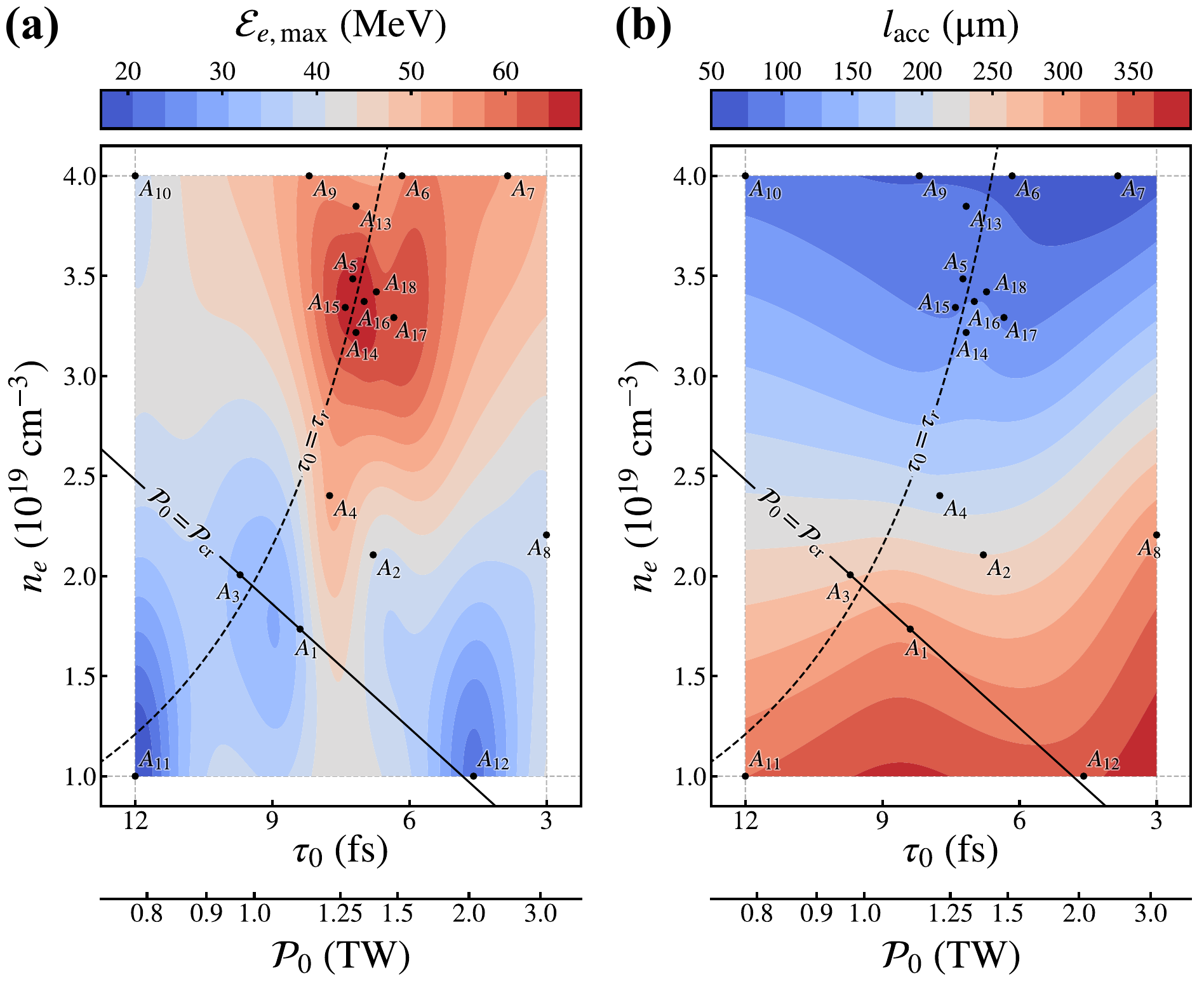}
\caption{(Color online). Self-guiding in a uniform-density plasma. (a) The GP model for the electron beam cut-off energy, $ \mathcal{E}_{e, \mathrm{max}} $, and (b) the corresponding acceleration distance, $ l_{\mathrm{acc}} $, based on the results of 18 PIC simulations, $ A_{1} $ -- $ A_{18} $ (parameters are listed in Tab.~\ref{tab:1}). The solid and dashed lines in both panels correspond to $ \mathcal{P}_0 = \mathcal{P}_{\mathrm{cr}} $ and $ \tau_0 = \tau_r $, respectively, as defined by Eqs.~(\ref{eq:p_0_over_p_cr_contour}) and (\ref{eq:tau_0_over_tau_r_contour}).}
\label{fig:2}
\end{figure}

First, we present the optimization results for the case of laser self-guiding in a uniform-density plasma. The GP model for $ \mathcal{E}_{e, \mathrm{max}} $ and the corresponding $ l_{\mathrm{acc}} $, based on the results of 18 PIC simulations (with parameters listed in Tab.~\ref{tab:1}), are shown in panels (a) and (b) of Fig.~\ref{fig:2}, respectively. The maximum cut-off energy of the electron beam, as estimated by the BO algorithm, is $ \mathcal{E}_{e, \mathrm{max}} \approx 68 \, \mathrm{MeV} $, and this energy is reached over an acceleration distance of $ l_{\mathrm{acc}} \approx 103 \, \upmu\mathrm{m} $. The optimal case is characterized by the following set of laser and plasma parameters: $ n_e \approx 3.36 \times 10^{19} \, \mathrm{cm^{-3}} $, $ \tau_0 \approx 7 \, \mathrm{fs} $, $ w_0 \approx 3 \, \upmu\mathrm{m} $, and $ a_0 \approx 2.65 $.

\begin{figure}[t]
\includegraphics[width=1.0\linewidth]{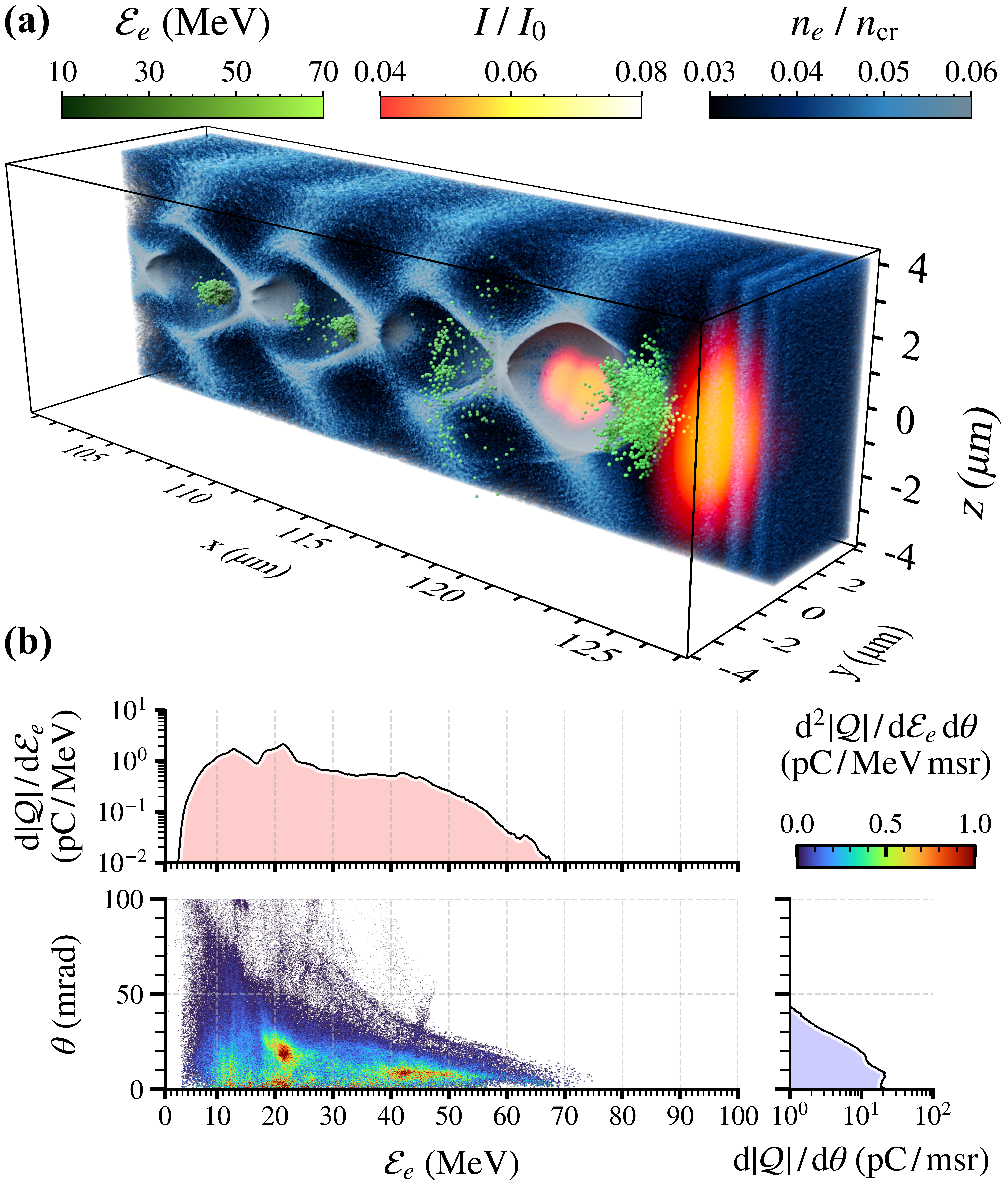}
\caption{(Color online). Spatial and spectral characteristics of the output electron beam obtained from $ A_{16} $ -- the PIC simulation with parameters closest to the optimum found by BO in the case of self-guiding in a uniform-density plasma (parameters are listed in Tab.~\ref{tab:1}). (a) 3D visualization of electron density, $ n_e / n_{\mathrm{cr}} $, laser intensity, $ I / I_0 $ (with $ I_0 = 2 \mathcal{P}_0 / \pi w_0^2 $), and the energy of accelerated electrons, $ \mathcal{E}_e $, considering only electrons with $ \mathcal{E}_e > 10 \, \mathrm{MeV} $. The electron density is sliced along the $ x $ -- $ z $ plane to reveal the inner structure of the plasma wave. (b) (top) Energy spectrum of the electron beam, $ \mathrm{d} |\mathcal{Q}| / \mathrm{d} \mathcal{E}_e $. (bottom left) Distribution of the electron charge with respect to energy and propagation angle, $ \mathrm{d}^2 |\mathcal{Q}| / \mathrm{d} \mathcal{E}_e \mathrm{d} \theta $. (bottom right) Angular spectrum of the electron beam, $ \mathrm{d} |\mathcal{Q}| / \mathrm{d} \theta $.}
\label{fig:3}
\end{figure}

As can be seen in panels (a) and (b) of Fig.~\ref{fig:2}, even within a relatively narrow parameter space, the cut-off energies and acceleration lengths vary significantly ($ \mathcal{E}_{e, \mathrm{max}} $ from $ \approx 20 $ to $ 70 \, \mathrm{MeV} $ and $ l_{\mathrm{acc}} $ from $ \approx 100 $ to $ 500 \, \upmu\mathrm{m} $), highlighting the importance of a proper choice of initial laser and plasma parameters. While the maximum cut-off energies are concentrated in a clearly identified region, the acceleration lengths increase primarily with decreasing electron density. It is important to note that in Fig.~\ref{fig:2}, not only the plasma density and pulse duration are varied, but also the laser strength parameter and beam waist are adjusted accordingly to ensure operation within the matched regime of LWFA.

As discussed in Sec.~\ref{sec:2}, in a uniform-density plasma, the laser pulse can be regarded as guided as long as the condition $ \mathcal{P}_0 \geq \mathcal{P}_{\mathrm{cr}} $ is satisfied. However, for a more precise description, the evolution of the laser during its interaction with the plasma must be considered, as this leads to laser energy depletion and the consequent loss of laser power. This power loss is partially compensated by plasma wave-induced compression of the laser pulse \cite{faure2005, schreiber2010}, which prolongs self-guiding. Nevertheless, once the power drops below the critical threshold for relativistic self-focusing, the laser pulse diffracts rapidly and electron acceleration ceases. This explains why, in panel (a) of Fig.~\ref{fig:2}, the region of maximum cut-off energies is located slightly above the contour curve $ \mathcal{P}_0 = \mathcal{P}_{\mathrm{cr}} $.

Panels (a) and (b) of Fig.~\ref{fig:3} display the spatial and spectral characteristics, respectively, of the output electron beam obtained from $ A_{16} $ -- the PIC simulation with parameters closest to the optimum found by BO in a uniform-density plasma (parameters are listed in Tab.~\ref{tab:1}). As can be seen, the energy spectrum is relatively broad, as it includes electrons injected into multiple periods of the plasma wave, not only via nanoparticle-assisted injection but also through self-injection occurring later during the interaction. The total charge of electrons with energy above $ 1 \, \mathrm{MeV} $ is $ \approx 40 \, \mathrm{pC} $, their angular distribution shows a FWHM divergence of $ \approx 37 \, \mathrm{mrad} $, and the efficiency of energy transfer from the laser to the accelerated electron beam is $ \approx 9.8 \, \% $. We note that these beam characteristics were not the subject of optimization.

\begin{figure}[t]
\includegraphics[width=1.0\linewidth]{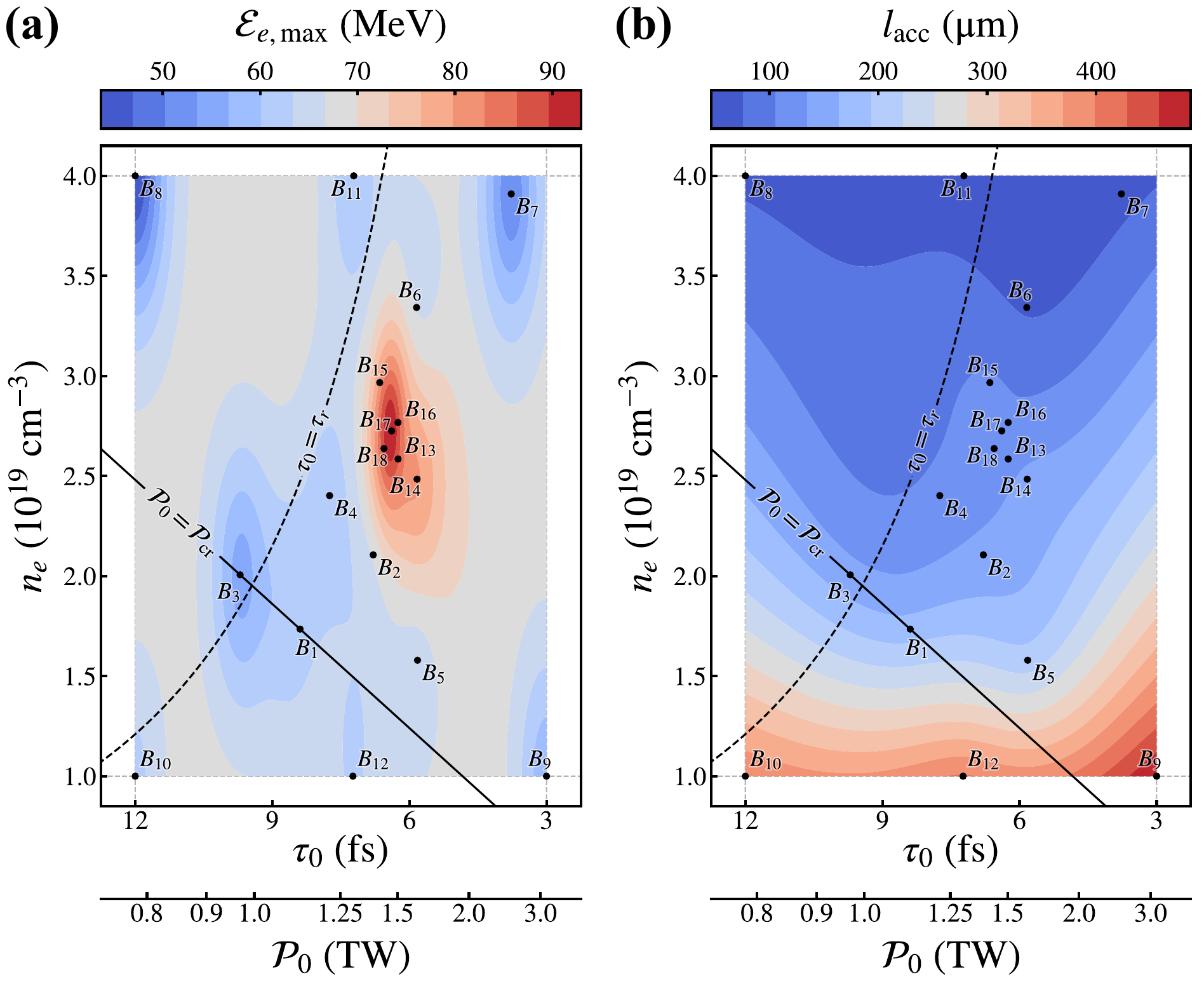}
\caption{(Color online). Guiding in a preformed plasma channel. (a) The GP model for the electron beam cut-off energy, $ \mathcal{E}_{e, \mathrm{max}} $, and (b) the corresponding acceleration distance, $ l_{\mathrm{acc}} $, based on the results of 18 PIC simulations, $ B_{1} $ -- $ B_{18} $ (parameters are listed in Tab.~\ref{tab:2}). The solid and dashed lines in both panels correspond to $ \mathcal{P}_0 = \mathcal{P}_{\mathrm{cr}} $ and $ \tau_0 = \tau_r $, respectively, as defined by Eqs.~(\ref{eq:p_0_over_p_cr_contour}) and (\ref{eq:tau_0_over_tau_r_contour}).}
\label{fig:4}
\end{figure}

\begin{figure}[t]
\includegraphics[width=1.0\linewidth]{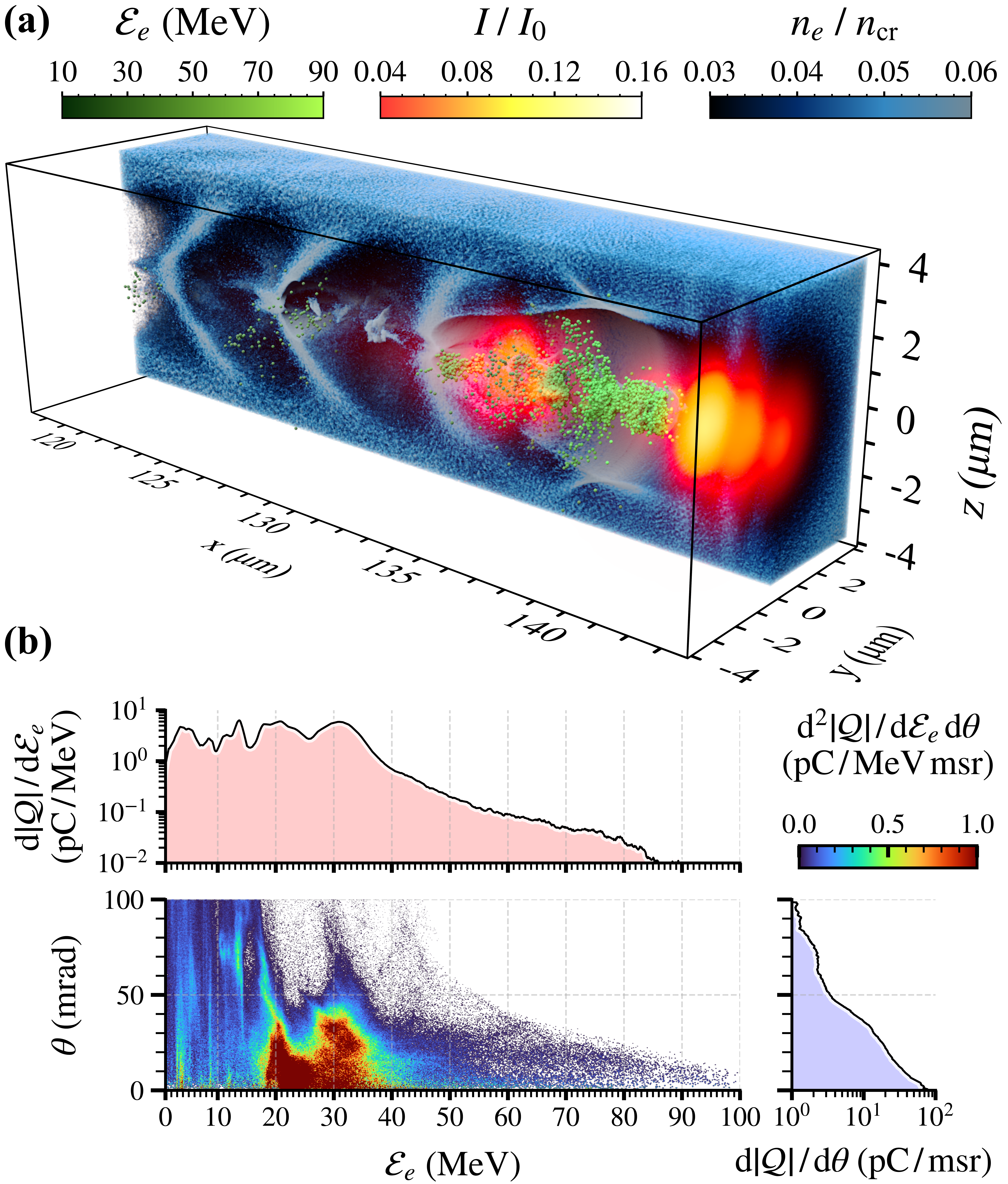}
\caption{(Color online). Spatial and spectral characteristics of the output electron beam obtained from $ B_{17} $ -- the PIC simulation with parameters closest to the optimum found by BO in the case of a preformed plasma channel with matched radius (parameters are listed in Tab.~\ref{tab:2}).  (a) The 3D visualization of electron density, $ n_e / n_{\mathrm{cr}} $, laser intensity, $ I / I_0 $, where $ I_0 = 2 \mathcal{P}_0 / \pi w_0^2 $, and the energy of accelerated electrons, $ \mathcal{E}_e $, considering only electrons with $ \mathcal{E}_e > 10 \, \mathrm{MeV} $. The electron density is sliced along the $ x $ -- $ z $ plane to reveal the inner structure of the plasma wave. (b) (top) The energy spectrum of the electron beam, $ \mathrm{d} \left| \mathcal{Q} \right| / \mathrm{d} \mathcal{E}_e $. (bottom left) The distribution of the electron charge with respect to the electron energy and propagation angle, $ \mathrm{d}^2 \left| \mathcal{Q} \right| / \mathrm{d} \mathcal{E}_e \mathrm{d} \theta $. (bottom right) The angular spectrum of the electron beam, $ \mathrm{d} \left| \mathcal{Q} \right| / \mathrm{d} \mathcal{\theta} $.}
\label{fig:5}
\end{figure}

The maximum electron energy according to Eq.~(\ref{eq:max_ene_guided}) for the optimal parameters found by BO is $ \mathcal{E}_{e, \mathrm{max}} \approx 62 \, \mathrm{MeV} $. The corresponding acceleration distance, determined by Eq.~(\ref{eq:l_acc_guided}), is $ l_{\mathrm{acc}} \approx 101 \, \upmu\mathrm{m} $. Both values are in good agreement with the simulation results. The slightly lower value of $ \mathcal{E}_{e, \mathrm{max}} $ predicted by Eq.~(\ref{eq:max_ene_guided}) may arise, among other factors, from the use of a linear approximation in calculating the accelerating field.

\subsection{Guiding in preformed plasma channel with matched radius \label{sec:6b}}

Second, we present the optimization results for the case of laser guiding in a preformed plasma channel with matched radius. The GP model for $ \mathcal{E}_{e, \mathrm{max}} $ and $ l_{\mathrm{acc}} $, based on the results of 18 PIC simulations (with parameters listed in Tab.~\ref{tab:2}), is shown in panels (a) and (b) of Fig.~\ref{fig:4}, respectively. The maximum cut-off energy of the electron beam, as estimated by the BO algorithm, is $ \mathcal{E}_{e, \mathrm{max}} \approx 93 \, \mathrm{MeV} $, and this energy is reached over an acceleration distance of $ l_{\mathrm{acc}} \approx 122 \, \upmu\mathrm{m} $. The optimal case is characterized by the following set of laser and plasma parameters: $ n_e \approx 2.73 \times 10^{19} \, \mathrm{cm^{-3}} $, $ \tau_0 \approx 6.4 \, \mathrm{fs} $, $ w_0 \approx 3.2 \, \upmu\mathrm{m} $, $ a_0 \approx 2.54 $, and $ R_{\mathrm{ch}} \approx 3 \, \upmu\mathrm{m} $.

As shown in panel (a) of Fig.~\ref{fig:4}, the highest electron cut-off energies are concentrated within a narrow region of the parameter space. Outside this region, the cut-off energy distribution becomes more uniform and approaches the maximum value observed for laser self-guiding in a uniform-density plasma. This illustrates the effect of a preformed plasma channel with matched radius, which suppresses laser diffraction and enables electron acceleration until the laser energy is significantly depleted or electron dephasing occurs. Relaxing the guiding conditions allows LWFA operation even when $ \mathcal{P}_0 / \mathcal{P}_{\mathrm{cr}} < 1 $, where the quality of the electron beam may improve, albeit at the cost of requiring a longer acceleration distance. As shown in panel (b) of Fig.~\ref{fig:4}, the acceleration distances again increase with decreasing electron density.

Panels (a) and (b) of Fig.~\ref{fig:5} show the spatial and spectral characteristics, respectively, of the output electron beam obtained from $ B_{17} $ -- the PIC simulation with parameters closest to the optimum found by BO in a preformed plasma channel (parameters are listed in Tab.~\ref{tab:2}). The total charge of electrons with energy above $ 1 \, \mathrm{MeV} $ is now $ \approx 149 \, \mathrm{pC} $, nearly four times higher compared to the optimal case in a uniform-density plasma. The angular distribution of the electron beam shows a FWHM divergence of $ \approx 18 \, \mathrm{mrad} $ and the efficiency of energy transfer from the laser to the accelerated electron beam is $ \approx 31.9 \, \% $. Again, no effort was made to optimize these characteristics.

For the optimal parameters, the maximum electron energy according to Eq.~(\ref{eq:max_ene_guided}) is $ \mathcal{E}_{e, \mathrm{max}} \approx 70 \, \mathrm{MeV} $, and the corresponding acceleration distance according to Eq.~(\ref{eq:l_acc_guided}) is $ l_{\mathrm{acc}} \approx 113 \, \upmu\mathrm{m} $. The discrepancy between the theoretical prediction and simulation results can be attributed to the same factors as in the case of self-guiding in a uniform-density plasma. Furthermore, the increase in electron energy and acceleration distance may be related to the mechanism described in Ref.~\citenum{shaw2017}: at later stages of the interaction, there is a significant overlap between the accelerated electrons and the red-shifted field of the laser pulse [see Fig.~\ref{fig:5}(a)], which may lead to the onset of direct laser acceleration \cite{pukhov1999, babjak2024, valenta2024}. This additional mechanism could enhance the final energy gain of the accelerated electrons, complementing the primary process of LWFA.

\section{Discussion on scaling \label{sec:7}}

The optimization of LWFA driven by relatively low-energy laser pulses (as in this work) is of particular interest for present-day high-repetition-rate laser systems \cite{faure2019, lazzarini2024a}. The scalability of the optimal parameters may be influenced by various factors, such as carrier-envelope phase effects for very short pulses \cite{valenta2020, huijts2022}, ion motion for very long pulses \cite{farina2001, zhou2012}, single-particle effects due to the distance between individual electrons in very low-density plasma, or particle collisions in very high-density plasmas. Verifying the optimized scaling for high-energy lasers through simulation requires advanced techniques (e.g., field decomposition into azimuthal Fourier modes \cite{lifschitz2009a} and the use of a Lorentz-boosted frame \cite{vay2007}) to address the increased computational load associated with long acceleration distances. We plan to pursue such verification in future work.

As shown in Sec.~\ref{sec:2}, LWFA operated in the matched regime can be described using only two parameters, $ n_e $ and $ \tau_0 $; alternatively, these parameters can be expressed in dimensionless form as $ \mathcal{P}^{*} \coloneq \mathcal{P}_0 / \mathcal{P}_{\mathrm{cr}} $ and $ \tau^{*} \coloneq \tau_0 \omega_p $. In general, $ \mathcal{P}^{*} $ and $ \tau^{*} $ may be functions that depend weakly on the laser energy. Treating them as constant for now, and disregarding the effects mentioned in the previous paragraph, we may derive the scaling of optimal LWFA input parameters, i.e., plasma density, pulse duration (or power), amplitude, spot size, and channel radius, all expressed in terms of laser energy:
\begin{equation}\label{eq:ne_scaling}
\begin{aligned}
    \frac{n_e}{n_{\mathrm{cr}}} &= \left( \frac{2 \sqrt{\pi \ln{2}}}{\tau^{*}} \right)^{-2/3} \left( \mathcal{P}^{*} \right)^{2/3} \left( \frac{\mathcal{E}_0}{\overline{\mathcal{E}}} \right)^{-2/3} \\
    &\approx 0.49 \, \left( \tau^{*} \right)^{2/3} \left( \mathcal{P}^{*} \right)^{2/3} \left( \frac{\mathcal{E}_0}{\overline{\mathcal{E}}} \right)^{-2/3},
\end{aligned}
\end{equation}
\begin{equation}\label{eq:tau0_scaling}
\begin{aligned}
    \frac{\tau_0}{T_0} &= \sqrt{\frac{\ln{2}}{\pi}} \left( \frac{2 \sqrt{\pi \ln{2}}}{\tau^{*}} \right)^{-2/3} \left( \mathcal{P}^{*} \right)^{-1/3} \left( \frac{\mathcal{E}_0}{\overline{\mathcal{E}}} \right)^{1/3} \\
    &\approx 0.23 \, \left( \tau^{*} \right)^{2/3} \left( \mathcal{P}^{*} \right)^{-1/3} \left( \frac{\mathcal{E}_0}{\overline{\mathcal{E}}} \right)^{1/3},
\end{aligned}
\end{equation}
\begin{equation}\label{eq:p0_scaling}
\begin{aligned}
\frac{\mathcal{P}_0}{\overline{\mathcal{P}}} &= \left( \frac{2 \sqrt{\pi \ln{2} }}{\tau^{*}} \right)^{2/3} \left( \mathcal{P}^{*} \right)^{1/3} \left( \frac{\mathcal{E}_0}{\overline{\mathcal{E}}} \right)^{2/3} \\ &\approx 2.06 \, \left( \tau^{*} \right)^{-2/3} \left( \mathcal{P}^{*} \right)^{1/3} \left( \frac{\mathcal{E}_0}{\overline{\mathcal{E}}} \right)^{2/3},
\end{aligned}
\end{equation}
\begin{equation}\label{eq:a0_scaling}
    a_0 = 2 \left( \mathcal{P}^{*} \right)^{1/3},
\end{equation}
\begin{equation}\label{eq:w0_scaling}
\begin{aligned}
    \frac{w_0}{\lambda_0} &= \frac{\sqrt{2}}{\pi} \left( \frac{2 \sqrt{\pi \ln{2}}}{\tau^{*}} \right)^{1/3} \left( \mathcal{P}^{*} \right)^{-1/6} \left( \frac{\mathcal{E}_0}{\overline{\mathcal{E}}} \right)^{1/3} \\
    &\approx 0.65 \, \left( \tau^{*} \right)^{-1/3} \left( \mathcal{P}^{*} \right)^{-1/6} \left( \frac{\mathcal{E}_0}{\overline{\mathcal{E}}} \right)^{1/3}.
\end{aligned}
\end{equation}
\begin{equation}\label{eq:Rch_scaling}
\begin{aligned}
    \frac{R_{\mathrm{ch}}}{\lambda_0} &= \frac{1.14}{\pi} \left( \frac{2 \sqrt{\pi \ln{2}}}{\tau^{*}} \right)^{1/3} \left( \frac{\mathcal{E}_0}{\overline{\mathcal{E}}} \right)^{1/3} \\
    &\approx 0.52 \, \left( \tau^{*} \right)^{-1/3} \left( \frac{\mathcal{E}_0}{\overline{\mathcal{E}}} \right)^{1/3}.
\end{aligned}
\end{equation}
Here, $ T_0 = 2\pi / \omega_0 $ is the laser period. Note that the optimal value of $ a_0 $ is independent of the laser pulse energy.

Similarly, we may derive the scaling of LWFA output parameters, i.e., electron energy and acceleration length, with respect to $ \mathcal{E}_0 $:
\begin{equation}\label{eq:ene_max_scaling}
\begin{aligned}
    \frac{\mathcal{E}_{e, \mathrm{max}}}{m_e c^2} &\approx 2.08 \, \left( \frac{2 \sqrt{\pi \ln{2}}}{\tau^{*}} \right)^{2/3} \left( \frac{\mathcal{E}_0}{\overline{\mathcal{E}}} \right)^{2/3} \\
    &\approx 4.28 \, \left( \tau^{*} \right)^{-2/3} \left( \frac{\mathcal{E}_0}{\overline{\mathcal{E}}} \right)^{2/3},
\end{aligned}
\end{equation}
\begin{equation}\label{eq:l_acc_scaling}
    \frac{l_{\mathrm{acc}}}{\lambda_0} \approx \sqrt{\frac{2 \ln{2}}{\pi}} \left( \mathcal{P}^{*} \right)^{-1} \frac{\mathcal{E}_0}{\overline{\mathcal{E}}} \approx 0.66 \, \left( \mathcal{P}^{*} \right)^{-1} \frac{\mathcal{E}_0}{\overline{\mathcal{E}}}.
\end{equation}
Interestingly, fitting of data from more than 50 published LWFA experiments reveals a scaling of electron energy with laser energy very close to a power of $ 2/3 $ as well \cite{labun2025}.

\begin{figure}[t]
\includegraphics[width=1.0\linewidth]{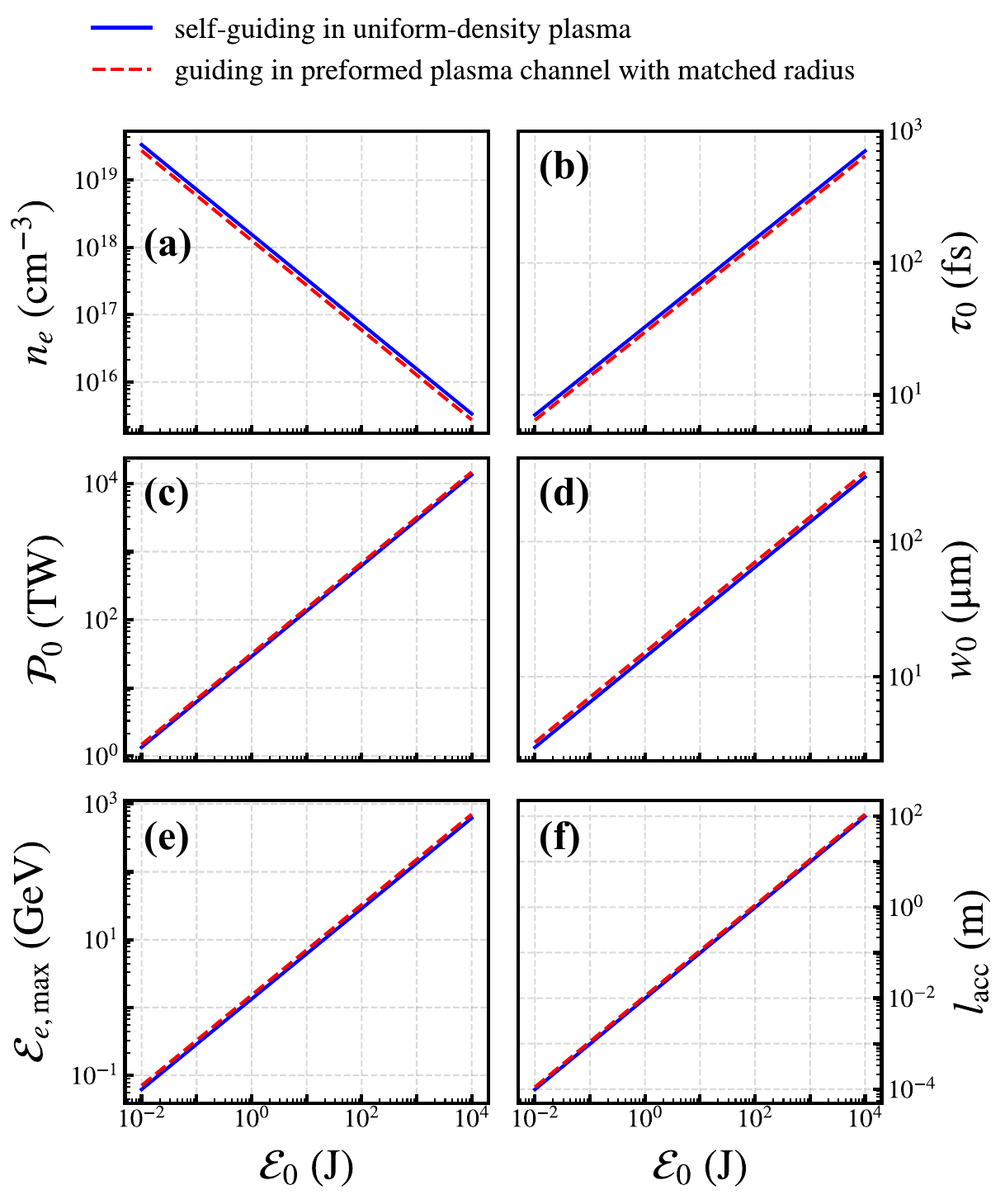}
\caption{(Color online). Scaling of optimal LWFA input parameters: (a) plasma density $ n_e $, (b) laser pulse duration $ \tau_0 $, (c) laser power $ \mathcal{P}_0 $, and (d) beam waist $ w_0 $; as well as output parameters: (e) electron beam cut-off energy $ \mathcal{E}_{e, \mathrm{max}} $, and (f) acceleration length $ l_{\mathrm{acc}} $, all expressed as functions of laser energy $ \mathcal{E}_0 $, according to Eqs.~(\ref{eq:ne_scaling}), (\ref{eq:tau0_scaling}), (\ref{eq:p0_scaling}), (\ref{eq:w0_scaling}), (\ref{eq:ene_max_scaling}), and (\ref{eq:l_acc_scaling}), respectively. The case of laser self-guiding in a uniform-density plasma (i.e., $ \mathcal{P}^{*} \approx 2.31 $ and $ \tau^{*} \approx 2.28 $) is shown by the solid line, and the case of laser guiding in a preformed plasma channel with matched radius (i.e., $ \mathcal{P}^{*} \approx 2.06 $ and $ \tau^{*} \approx 1.88 $) is shown by the dashed line.}
\label{fig:6}
\end{figure}

The number of electrons that can be injected into the wakefield for acceleration before LWFA becomes severely loaded was found to be $ \mathcal{N}_{e} \approx \lambda_0 \sqrt{ \mathcal{P}_0 / \overline{\mathcal{P}} } / 6 \pi r_e $ \cite{lu2007, bulanov2016}. When rewritten in terms of $\mathcal{E}_0$, this expression reads as
\begin{equation}\label{eq:n_e_scaling}
\begin{aligned}
    \mathcal{N}_{e} &\approx \frac{\lambda_0}{6 \pi r_e} \left( \frac{2 \sqrt{\pi \ln 2}}{\tau^{*}} \right)^{1/3} \left( \mathcal{P}^{*} \right)^{1/6} \left( \frac{\mathcal{E}_0}{\overline{\mathcal{E}}} \right)^{1/3} \\
    &\approx 0.08 \, \frac{\lambda_0}{r_e} \left( \tau^{*} \right)^{-1/3} \left( \mathcal{P}^{*} \right)^{1/6} \left( \frac{\mathcal{E}_0}{\overline{\mathcal{E}}} \right)^{1/3}.
\end{aligned}
\end{equation}
We see that the upper limit of the efficiency of energy transfer from the laser to the accelerated electron beam depends on neither the laser energy nor the wavelength,
\begin{equation}\label{eq:efficiency_scaling}
    \eta = \frac{\mathcal{N}_{e} \mathcal{E}_{e, \mathrm{max}}}{\mathcal{E}_{0}} \approx 0.34 \, \left( \tau^{*} \right)^{-1} \left( \mathcal{P}^{*} \right)^{1/6}.
\end{equation}

In the case of laser self-guiding in a uniform-density plasma (Sec.~\ref{sec:6a}), the optimal case found in this work corresponds to $ \mathcal{P}^{*} \approx 2.31 $ and $ \tau^{*} \approx 2.28 $. Using Eqs.~(\ref{eq:ne_scaling}), (\ref{eq:tau0_scaling}), (\ref{eq:a0_scaling}), and (\ref{eq:w0_scaling}) to rescale the optimal parameters to, e.g., a $ 1 \, \mathrm{kJ} $ laser pulse, we obtain $ n_e \approx 1.56 \times 10^{16} \, \mathrm{cm^{-3}} $, $ \tau_0 \approx 327 \, \mathrm{fs} $, $ a_0 \approx 2.65 $, and $ w_0 \approx 140 \, \upmu\mathrm{m} $. For such a hypothetical LWFA setup, Eqs.~(\ref{eq:ene_max_scaling})--(\ref{eq:efficiency_scaling}) predict a maximum electron energy of $ \approx 134 \, \mathrm{GeV}$, an acceleration length of $ \approx 9.8 \, \mathrm{m} $, an electron charge of $ \approx 1.2 \, \mathrm{nC} $, and an acceleration efficiency of $ 17 \, \% $, respectively.

In the case of laser guiding in a preformed plasma channel with matched radius (Sec.~\ref{sec:6b}), the optimal case corresponds to $ \mathcal{P}^{*} \approx 2.06 $ and $ \tau^{*} \approx 1.88 $. Considering again a $ 1 \, \mathrm{kJ} $ laser pulse, the optimal parameters translate to $ n_e \approx 1.27 \times 10^{16} \, \mathrm{cm^{-3}} $, $ \tau_0 \approx 299 \, \mathrm{fs} $, $ a_0 \approx 2.54 $, $ w_0 \approx 152 \, \upmu\mathrm{m} $, and $ R_{\mathrm{ch}} \approx 137 \, \upmu\mathrm{m} $. Such a setup would yield an electron beam with a cut-off energy of $ \approx 161 \, \mathrm{GeV} $ over an acceleration distance of $ \approx 11 \, \mathrm{m} $, a charge of $ \approx 1.3 \, \mathrm{nC} $, and an efficiency of $ 20 \, \% $.

Panels (a)--(f) of Fig.~\ref{fig:6} show the scaling of optimal plasma density, laser pulse duration, laser power, beam waist, electron beam cut-off energy, and acceleration length, respectively, all expressed as functions of laser energy, according to Eqs.~(\ref{eq:ne_scaling}), (\ref{eq:tau0_scaling}), (\ref{eq:p0_scaling}), (\ref{eq:w0_scaling}), (\ref{eq:ene_max_scaling}), and (\ref{eq:l_acc_scaling}). Results are shown for both cases: laser self-guiding in a uniform-density plasma and guiding in a preformed plasma channel with matched radius.

\section{Conclusion \label{sec:8}}

By leveraging the BO method and 3D PIC simulations, we efficiently identify the LWFA regime that maximizes the cut-off energy of an electron beam while minimizing the number of simulations required. We show that, with a $ 10 \, \mathrm{mJ} $ laser pulse and appropriately chosen laser and plasma parameters, LWFA can produce electron beams with cut-off energies approaching $ 70 \, \mathrm{MeV} $ using self-guiding in a uniform-density plasma, and exceeding $ 90 \, \mathrm{MeV} $ when guided in a preformed plasma channel with matched radius, both achieved over acceleration distances slightly above $ 100 \, \upmu\mathrm{m} $.

To interpret the simulation results quantitatively, we derive novel analytical expressions for predicting the maximum electron energy and the corresponding acceleration length, accounting for the effects of laser pulse diffraction and energy depletion. The model shows relatively good agreement with the simulation results.

Additionally, we express the optimization results in terms of dimensionless parameters, which may aid in developing generalized scaling laws for LWFA driven by lasers of arbitrary energy. As a step in this direction, we outline how the scaling of electron energy and the corresponding acceleration length could be formulated, i.e., expressed as functions of laser energy, optimized to yield the maximum possible electron energy for a given laser energy, and accompanied by the full set of input parameters required to achieve such scaling.

\section*{Acknowledgements \label{sec:9}}

We acknowledge fruitful discussions with S.~S.~Bulanov, G.~M.~Grittani, M.~Jech, M.~Kando, K.~G.~Miller, A.~S.~Pirozhkov, B.~A.~Reagan, B.~Rus, B.~K.~Russell, and P.~V.~Sasorov. 

This work was supported by the Defense Advanced Research Program Agency (DARPA) under the Muons for Science and Security Program and by the NSF and Czech Science Foundation (NSF-GACR collaborative Grant No. 2206059 and Czech Science Foundation Grant No. 22-42963L). This work was supported by the project ``e-INFRA CZ'' (ID:90254) from the Ministry of Education, Youth and Sports of the Czech Republic. A portion of this work was performed under the auspices of the U.S. Department of Energy by Lawrence Livermore National Laboratory (LLNL) under Contract DE-AC52-07NA27344 and supported by the LLNL Institutional Computing Grand Challenge program. The EPOCH code used in this work was in part funded by the UK EPSRC grants EP/G054950/1, EP/G056803/1, EP/G055165/1, EP/M022463/1, and EP/P02212X/1. We acknowledge ``Bayesian Optimization: Open source constrained global optimization tool for Python'' \cite{nogueira2014}.

\section*{Data availability \label{sec:10}}

The data that support the findings of this article are openly available \cite{valenta2025a}.

\appendix*
\section{Simulation parameters \label{sec:11}}

Tabs.~\ref{tab:1} and \ref{tab:2} contain the parameters of the PIC simulations as well as the corresponding cut-off electron energy and acceleration length for the cases of laser self-guiding in a uniform-density plasma and guiding in a preformed plasma channel with matched radius, respectively.

\bgroup
\def\arraystretch{1.2}%
\begin{table*}[h!]
\begin{tabular}{ccccccc}
\hline \hline
No. & $ n_e \left( 10^{19} \mathrm{cm^{-3}} \right) $ & $ \tau_0 \left( \mathrm{fs} \right) $ & $ w_0 \left( \mathrm{\upmu m} \right) $ & $ a_0 $ & $ \mathcal{E}_{e, \mathrm{max}} \left( \mathrm{MeV} \right) $ & $ l_{\mathrm{acc}} \left( \mathrm{\upmu m} \right) $ \\ \hline
$ A_1 $ & $ 1.74 $ & 8.39 & 3.61 & 2.00 & 32.6 & 299  \\
$ A_2 $ & $ 2.11 $ & 6.79 & 3.50 & 2.29 & 39.7 & 241  \\
$ A_3 $ & $ 2.01 $ & 9.71 & 3.35 & 2.00 & 33.9 & 249  \\
$ A_4 $ & $ 2.40 $ & 7.75 & 3.28 & 2.29 & 50.8 & 188  \\
$ A_5 $ & $ 3.49 $ & 7.24 & 2.93 & 2.65 & 66.2 & 102  \\
$ A_6 $ & $ 4.00 $ & 6.16 & 2.87 & 2.93 & 57.2 & 65  \\
$ A_7 $ & $ 4.00 $ & 3.85 & 3.11 & 3.43 & 54.2 & 52  \\
$ A_8 $ & $ 2.21 $ & 3.00 & 3.95 & 3.05 & 39.3 & 308  \\
$ A_{9} $ & $ 4.00 $ & 8.19 & 2.74 & 2.66 & 55.6 & 74  \\
$ A_{10} $ & $ 4.00 $ & 12.00 & 2.57 & 2.35 & 40.1 & 89  \\
$ A_{11} $ & $ 1.00 $ & 12.00 & 4.08 & 1.48 & 17.2 & 338  \\
$ A_{12} $ & $ 1.00 $ & 4.60 & 4.79 & 2.04 & 22.8 & 357  \\
$ A_{13} $ & $ 3.85 $ & 7.17 & 2.84 & 2.75 & 57.6 & 82  \\
$ A_{14} $ & $ 3.22 $ & 7.17 & 3.01 & 2.59 & 66.0 & 120  \\
$ A_{15} $ & $ 3.34 $ & 7.40 & 2.96 & 2.60 & 66.0 & 98  \\
$ A_{16} $ & $ 3.37 $ & 6.99 & 2.98 & 2.65 & 67.9 & 102  \\
$ A_{17} $ & $ 3.29 $ & 6.34 & 3.05 & 2.72 & 63.9 & 97  \\
$ A_{18} $ & $ 3.42 $ & 6.72 & 2.99 & 2.70 & 63.9 & 100  \\
\hline \hline
\end{tabular}
\caption{\label{tab:1} Parameters of PIC simulations with a uniform-density plasma. Electron density, $ n_e $, pulse duration, $ \tau_0 $, beam waist, $ w_0 $, and strength parameter, $ a_0 $, prescribed in PIC simulations $ A_{1} $ - $ A_{18} $ as well as the corresponding cut-off electron energy $ \mathcal{E}_{e, \mathrm{max}} $ acquired over acceleration distance $ l_{\mathrm{acc}} $.}
\end{table*}
\egroup

\bgroup
\def\arraystretch{1.2}%
\begin{table*}[h!]
\begin{tabular}{cccccccc}
\hline \hline
No. & $ n_e \left( 10^{19} \mathrm{cm^{-3}} \right) $ & $ \tau_0 \left( \mathrm{fs} \right) $ & $ w_0 \left( \mathrm{\upmu m} \right) $ & $ a_0 $ & $ R_{\mathrm{ch}} \left( \mathrm{\upmu m} \right) $ & $ \mathcal{E}_{e, \mathrm{max}} \left( \mathrm{MeV} \right) $ & $ l_{\mathrm{acc}} \left( \mathrm{\upmu m} \right) $ \\ \hline
$ B_1 $ & $ 1.74 $ & 8.39 & 3.61 & 2.00 & 2.94 & 60.4 & 168 \\
$ B_2 $ & $ 2.11 $ & 6.79 & 3.50 & 2.29 & 3.05 & 67.3 & 145 \\
$ B_3 $ & $ 2.01 $ & 9.71 & 3.35 & 2.00 & 2.73 & 55.3 & 146 \\
$ B_4 $ & $ 2.40 $ & 7.75 & 3.28 & 2.29 & 2.86 & 61.4 & 113 \\
$ B_5 $ & $ 1.58 $ & 5.82 & 3.96 & 2.19 & 3.37 & 64.8 & 208 \\
$ B_6 $ & $ 3.34 $ & 5.84 & 3.08 & 2.81 & 2.97 & 66.5 & 73 \\
$ B_7 $ & $ 3.91 $ & 3.77 & 3.14 & 3.42 & 3.35 & 51.0 & 66 \\
$ B_8 $ & $ 4.00 $ & 12.00 & 2.57 & 2.35 & 2.27 & 43.7 & 69 \\
$ B_9 $ & $ 1.00 $ & 3.00 & 5.15 & 2.35 & 4.54 & 57.8 & 487 \\
$ B_{10} $ & $ 1.00 $ & 12.00 & 4.08 & 1.48 & 2.86 & 62.5 & 397 \\
$ B_{11} $ & $ 4.00 $ & 7.22 & 2.80 & 2.78 & 2.69 & 60.6 & 59 \\
$ B_{12} $ & $ 1.00 $ & 7.24 & 4.44 & 1.75 & 3.38 & 62.9 & 402  \\
$ B_{13} $ & $ 2.58 $ & 6.25 & 3.32 & 2.52 & 3.03 & 87.0 & 126  \\
$ B_{14} $ & $ 2.48 $ & 5.83 & 3.40 & 2.55 & 3.13 & 80.3 & 140  \\
$ B_{15} $ & $ 2.97 $ & 6.65 & 3.14 & 2.59 & 2.90 & 80.6 & 109  \\
$ B_{16} $ & $ 2.77 $ & 6.25 & 3.24 & 2.58 & 3.00 & 88.3 & 115  \\
$ B_{17} $ & $ 2.73 $ & 6.39 & 3.25 & 2.55 & 2.99 & 92.8 & 122  \\
$ B_{18} $ & $ 2.64 $ & 6.56 & 3.27 & 2.50 & 2.98 & 88.9 & 118  \\
\hline \hline
\end{tabular}
\caption{\label{tab:2} Parameters of PIC simulations with a preformed plasma channel of matched radius. Electron density, $ n_e $, pulse duration, $ \tau_0 $, beam waist, $ w_0 $, strength parameter, $ a_0 $, and the channel radius, $ R_{\mathrm{ch}} $, prescribed in PIC simulations $ B_{1} $ - $ B_{18} $ as well as the corresponding cut-off electron energy $ \mathcal{E}_{e, \mathrm{max}} $ acquired over acceleration distance $ l_{\mathrm{acc}} $.}
\end{table*}
\egroup

\end{document}